\documentclass[mathleft,fleqn,% %final,%%earlyview,%
]{an}
\usepackage{amsmath}
\usepackage{array}
\usepackage{graphicx}
\usepackage[varg]{txfonts}
\usepackage{natbib}
\bibpunct{(}{)}{;}{a}{}{,}
\begin{document}

% The following seven commands are intended for editorial usage and
% should be ignored by the author(s).
\Pagespan{1}{}% Document's page range.
% If second parameter is left empty, the last page is computed
% automatically.
\Yearpublication{2016}%
\Yearsubmission{2016}%
\Month{0}%
\Volume{999}%
\Issue{0}%
\DOI{asna.201400000}%

\title{Magnetic field gradient effects on the magnetorotational instability}

\author{Suzan Do\u{g}an\inst{1,2}\fnmsep\thanks{e-mail:
        {suzan.dogan@ege.edu.tr}}}
% Example for footnote, note the usage of the \texttt{fnmsep} command
% as separator between institute number and footnote mark}

\titlerunning{Magnetic field gradient effects on MRI}
\authorrunning{S. Do\u{g}an}
\institute{
Department of Astronomy and Space Sciences, Faculty of Science, University of Ege, Bornova, 35100, \.Izmir, Turkey.
\and
Department of Physics and Astronomy, University of Leicester, University Road, Leicester LE1 7RH, UK.}

\received{XXXX}
\accepted{XXXX}
\publonline{XXXX}


\abstract{The magnetorotational instability (MRI), also known as the Balbus -- Hawley instability, is thought to have an important role on the initiation of turbulence and angular momentum transport in accretion discs. In this work, we investigate the effect of the magnetic field gradient in the azimuthal direction on MRI. We solve the magnetohydrodynamic equations by including the azimuthal component of the field gradient. We find the dispersion relation and calculate the growth rates of the instability numerically. The inclusion of the azimuthal magnetic field gradient produces a new unstable region on wavenumber space. It also modifies the growth rate and the wavelength range of the unstable mode: the higher the magnitude of the field gradient, the greater the growth rate and the wider the unstable wavenumber range. Such a gradient in the magnetic field may be important in T Tauri discs where the stellar magnetic field has an axis which is misaligned with respect to the rotation axis of the disc.}

\maketitle

\section{Introduction}
An accretion disc with angular velocity decreasing radially outward and threaded by a weak magnetic field is dynamically unstable \citep{BH1991}. This unstable behavior, which is known as the magnetorotational instability (MRI), rapidly generates magnetohydrodynamic (MHD) turbulence which is probably the origin of \textquotedblleft anomalous viscosity\textquotedblright \, in accretion discs. MHD turbulence driven by MRI effectively transports angular momentum radially outward \citep{BH1998}.

After \cite{BH1991} established the importance of MRI in the structure and the dynamics of the accretion discs, it was pointed out by many authors that the non-ideal MHD effects play a significant role in modifying the MRI. Non-ideal MHD effects must be considered in systems where the accretion discs are weakly ionized. For instance, protoplanetary discs (PPDs) are too dense and cool, so the thermal ionization process is expected to be ineffective. Low ionization fraction makes the non-ideal MHD effects important in such discs, as the plasma is not well-coupled to the magnetic field. The outer regions of dwarf novae discs may also be weakly ionized and the non-ideal MHD effects are again expected to be important (\citealt{Gammie1996}; \citealt{GM1998}; \citealt{Stoneetal2000}).

The importance of the Hall effect in modifying the maximum growth rate and the characteristic wavelength of MRI was first pointed out by \cite{Wardle1999}. \cite{Wardle1999} found that the growth rate of the instability depends on the relative signs of the initially vertical magnetic field and the angular momentum of the disc, i.e. whether $\mathbf{\Omega} \cdot \textbf{\emph{B}}$ is positive or negative where $\mathbf{\Omega}$ is the angular velocity vector of the disc and $\textbf{\emph{B}}$ is the magnetic field vector. When the Hall current is dominated by the negative (positive) species, the parallel case ($\mathbf{\Omega} \cdot \textbf{\emph{B}} > 0$) becomes less (more) unstable than the anti-parallel case ($\mathbf{\Omega} \cdot \textbf{\emph{B}} < 0$). \cite{BT2001} (hereafter BT01) analysed the Hall effect in protostellar discs and found that the inclusion of the Hall effect destabilizes the disc with any differential rotation law. \cite{SS2002}, who investigated the effect of the Hall term on the evolution of MRI in weakly ionized discs, showed that the Hall term is important to the linear properties of the MRI and therefore must be included in models of both dwarf nova discs in quiescence, and in PPDs. Recent works (e.g., \citealt{WS2012}; \citealt{KL2013}; \citealt{Bai2014,Bai2015}) have also emphasized that the inclusion of Hall effect makes the properties of the MRI depend on the orientation of the net vertical field with respect to the disc rotation axis in PPDs.

In addition to local analyses, several global MRI calculations have been performed for discs and spheres over the last two decades. \cite{RK2005} considered the influence of the Hall effect on the global stability of cool protoplanetary discs. They determined the minimal and maximal magnetic field amplitudes and the values of critical Reynolds numbers for which MRI would occur. They showed that the MRI of a Kepler flow is significantly modified by the Hall effect depending on the orientation of the magnetic field in relation to the rotation axis. \cite{Kondicetal2011,Kondicetal2012} investigated the stability of the Hall-MHD system for a spherical shell and determined its importance for newborn neutron stars. Similarly, their results differ for magnetic fields aligned with the rotation axis and anti-parallel magnetic fields.

For simplicity, most theoretical models assume that the magnetic axis and the rotation axis of the disc/star coincide. This is, however, may not be the case in most of the astrophysical systems. Misalignment can occur in systems where the central object has a dipole field with a magnetic axis misaligned with respect to its spin axis (e.g. \citealt{Alencar2002}; \citealt{Symington2005}). We note that the magnetic field geometry of PPDs may not be purely dipole (\citealt{Donatietal2008}; \citealt{JKetal1999}; \citealt{ValentiJK2004}). Their structure is probably more complicated, however it is expected that the dipole component of the field dominates at distances of several stellar radii \citep{JK2007}. In any case, when the magnetic and the rotation axis are not exactly aligned, the disc particles \textquotedblleft feel\textquotedblright $\rm$ a magnetic field gradient in both radial and azimuthal directions as their distance from the magnetic poles and the equator changes during their motion per orbit. It has recently shown by \cite{DP2007} and \cite{DP2012} (hereafter DP12) that the magnetic field gradient produced in the radial direction changes the growth rates of the MRI significantly. The inclusion of the gradient terms produces new unstable regions and increases the maximum value of the growth rates. In the present investigation, we aim to examine the effect of the magnetic field gradient which is produced in the azimuthal direction in the disc. Such a gradient can be generated by the misalignment of the magnetic axis with respect to the rotation axis of the disc.

The plan of the paper is as follows: In Section 2, we describe the basic geometry of the problem, present the linearized MHD equations that include the terms describing the magnetic field gradient in the azimuthal direction and find the dispersion relation. In Section 3, the numerical growth rates found from the dispersion relation are presented both for zero and finite resistivity. The effect of the magnetic field gradient on the instability is investigated. We finally discuss our results in Section 4.

\section{Basics and Formulation}
\subsection{Misaligned dipole}
Let us begin by considering the simple geometry such that the disc fluid orbits a central object with a dipole field and where the stellar magnetic axis and rotation axis of the disc are aligned. If we use standard cylindrical coordinates ($R,\phi,z$) with the origin at the disc center, then the magnetic and the rotation axes coincide on the z-axis. Therefore, the magnetic equator lies in the disc plane and the magnetic field strength remains constant for the drifting fluid particles at the same radial distance. The particles feel no magnetic field gradient in the azimuthal direction during their orbital motion. This is the case where the configuration is axisymmetric.

When the magnetic axis and rotation axis of the disc are not aligned, the drifting particles spend half the \emph{drift period} on the northern magnetic hemisphere and the other half
on the southern one. In this case, both the magnetic latitude and the distance from the magnetic axis vary during their per orbital motion. Therefore, the particles encounter a magnetic field gradient in the azimuthal direction due to the fact that the magnetic field strength $B \propto (1+3\sin^2 \Lambda)^{1/2}/R^3$ where \emph{R} is the distance from the central star and $\Lambda$ is the magnetic latitude measured northwards from the magnetic
equator. When the particles move towards the region of the disc which is close to the magnetic pole (equator), the magnetic field strength increases (decreases). We know that the MRI occurs only if the magnetic field is weak (i.e. sub-thermal). Therefore, we expect to see the effect of the fluctuation of magnetic field on the instability. If the particles experience a decrease in an already weak field, the instability is expected to become more powerful. We should also mention that the MRI may switch off if the field is too weak. In this context, global calculations by \cite{RK2005} point out that if the external magnetic field strength does not exceed 0.1 G then the MRI can not occur without the Hall effect of parallel fields.

In a dipole field we know that the total field vector is purely vertical at the magnetic equator, i.e. \textbf{B}=$B_z\hat{z}$ at $\Lambda = 0^\circ$. Therefore, the magnetic field lines are perpendicular to the disc plane when the magnetic axis of the central object and the rotation axis of the disc are aligned. However, in a misaligned case the magnetic field threading the disc will also have radial and azimuthal components ($B_R,B_{\phi}$) in addition to the vertical component. The magnitude of each component in the disc plane differs from those they have in the aligned case. The importance of the R and $\phi$ components depend on the inclination angle of the magnetic axis. In this study, we consider the vertical component of the magnetic field ($B_z$) only, as it is crucial for MRI. Besides, the presence of a radial component causes a linear time dependence in $B_{\phi}$. Therefore, it is mathematically the simplest to consider the case of vanishing radial field \citep{BH1998}.

\subsection{MHD Equations}
The fundamental equations are mass conservation,
\begin{equation}
\frac{\partial \rho }{\partial t} +\nabla \cdot \left(\rho \,
\textbf{\emph{v}}\right)=0
\end{equation}
the equation of motion,
\begin{equation}
\rho \frac{\partial \textbf{\emph{v}}}{\partial t} +\left(\rho
\textbf{\emph{v}}\cdot \nabla \right)\textbf{\emph{v}}=-\nabla
P+\frac{1}{c} \textbf{\emph{J}}\times \textbf{\emph{B}}
\end{equation}
and the induction equation,
\begin{equation}
\frac{\partial \textbf{\emph{B}}}{\partial t} =\nabla \times
\left[\emph{\textbf{v}}\times \textbf{\emph{B}}-\eta \frac{4\pi }{c}
\textbf{\emph{J}}-\frac{\textbf{\emph{J}}\times
\textbf{\emph{B}}}{en_{e} } \right]
\end{equation}
where $\textbf{\emph{v}}$ is the fluid velocity, $\eta$ is the microscopic resistivity and $n_e$ is the number density of electrons. Here the current density (\textbf{\emph{J}}) is given as
\begin{equation}
\textbf{\emph{J}}=\frac{c}{4\pi}\nabla \times \textbf{\emph{B}}.
\end{equation}

In a misaligned dipole, the azimuthal gradient in magnetic field ($\nabla B$) is expected to generate a magnetic pressure force which pushes the plasma particles towards the negative  $\nabla B$ direction. If the frozen-in condition is satisfied, frozen-in particles bend the magnetic field lines in such a way as to give them a curvature. As a result, the curvature of the magnetic field lines produces a magnetic tension force in the opposite direction of the magnetic pressure force. Therefore, the magnetic tension force balances the magnetic pressure gradient in the equilibrium. In our stability analysis, we have ignored the current produced by the electrons under the influence of curvature drift. We use standard cylindrical coordinates ($R,\phi,z$) with the origin at the disc
center. We consider the local stability of a Keplerian disc threaded
by a vertical field with a gradient in the azimuthal direction,
$\emph{\textbf{B}} = B(\phi)\hat{z}$. Therefore the gradient
of the magnetic field is $\nabla B= (d B/Rd \phi)\hat{\mathbf{\phi}}$.
We shall work in the Boussinesq limit which is frequently used in
descriptions of the nature of accretion disc transport (e.g. \citealt{BH1991};
\citealt{BT2001}). We should note that the disc fluid is not exactly but nearly
incompressible \citep{BH1998}. Therefore, the velocity field in accretion discs may be
taken as ($\nabla\cdot\textbf{u}=0$) for the turbulent flows. Finally, we assume
that the perturbed quantities vary in space
and time like a plane wave, i.e., $\exp(ikz+\omega t)$, where
$\textbf{k}$ is the wave vector perpendicular to the disc and
$\omega$ is the angular frequency. We find the linearized radial, azimuthal and vertical components
of equation of motion as
\begin{equation}
\omega \delta v_{R}-2\Omega \delta v_{\phi}-\frac{ik }{4\pi \rho} B\delta
B_{R}=0
\end{equation}
\begin{equation}
\omega \delta v_{\phi } +\frac{\kappa ^{2} }{2\Omega } \delta v_{R}
-\frac{1 }{4\pi \rho } \left[ik B \delta B_{\phi }-\frac{\partial B}{R \partial \phi}\delta B_{z}\right ] =0
\end{equation}
\\
\begin{equation}
ik\frac{ \delta P}{\rho } -\frac{1}{4\pi \rho }\frac{\partial B}{R \partial \phi}
\delta B_{\phi} =0.
\end{equation}
Including the finite resistivity and the Hall effect, we find the same components of the linearized induction equation as
\begin{equation}
(k^2\eta+\omega)\delta B_{R}-ikB \delta v_{R}+\frac{c}{4\pi en_{e} } \left[k^{2}B  \delta B_{\phi }+ \frac{\partial B}{R \partial \phi} \delta B_{z}\right ]=0
\end{equation}
\\
\begin{equation}
(k^2\eta+\omega)\delta B_{\phi}-ikB \delta v_{\phi}-\left[\frac{d\Omega}{d ln R}+\frac{c}{4\pi en_{e} }B k^{2} \right]\delta B_{R }=0
\end{equation}
\\
\begin{equation}
(k^2\eta+\omega)\delta B_{z}+\frac{\partial B}{R \partial \phi}\delta v_{\phi}-ik\frac{c}{4\pi en_{e} }\frac{\partial B}{R \partial \phi}\delta B_{R }=0.
\end{equation}
Here $\kappa^2=4\Omega^{2}+d\Omega^{2}/d\ln R $ is the epicyclic
frequency and $\mathbf{\Omega}$ is the angular velocity of the disc. The Eqns. (5) - (10)  yield a fifth - order dispersion relation given below:
\begin{equation}
\begin{array}{l}
{\omega^5+3k^2 \eta \omega^4+ \omega^3 \left\{\begin{array}{l}{\kappa^2+k^2 v_A^2\left[\dfrac{3k^2\eta}{\Omega}+2-\dfrac{{G}^2} {k^2}\right]}\\
\\
{+\dfrac{k^2 v_H^2}{4\Omega} \left[\dfrac{\rm d \, \Omega^2}{\rm d \,\rm ln R} - \dfrac{k^2v_A^2v_H^2}{\Omega^2}{G}^2  + k^2v_H^2\right]}\end{array}\right\}}
\\
\\
{+\omega^2k^2\eta\left\{\begin{array}{l}{3{\kappa}^2+k^2v_A^2\left[\dfrac{k^2\eta^2}{v_A^2}+4-\dfrac{{2 G}^2} {k^2}\right]}\\
\\
{+\dfrac{k^2v_H^2}{4} \left[\dfrac{\rm d \, \Omega^2}{\rm d \,\rm ln R} - \dfrac{k^2v_H^2v_A^2}{\Omega^4}{G}^2  + \dfrac{k^2v_H^2}{\Omega^2} \right]}\end{array}\right\}}\\
\\
{+\omega\left\{\begin{array}{l}{\left[k^2v_A^2+\dfrac{k^2v_H^2}{4\Omega^2}\kappa^2 \right]\left[\dfrac{\rm d \, \Omega^2}{\rm d \,\rm ln R}+k^2\big(v_A^2 + v_H^2\big)\bigg(1-\dfrac{{G}^2} {k^2}\bigg)\right]}\\
\\
{+\dfrac{k^2\eta^2}{v_A^2}\left[\kappa^2+k^2v_A^2\bigg(2-\dfrac{{G}^2} {k^2}\bigg)\right]}\end{array}\right\}}\\
\\
{+\eta k^4\left( v_A^2+\dfrac{v_H^2\kappa^2}{4\Omega^2}\right)\left[\dfrac{\rm d \, \Omega^2}{\rm d \,\rm ln R}+k^2\big(v_A^2 + v_H^2\big)\bigg(1-\dfrac{{G}^2} {k^2}\bigg) \right] =0}
\end{array}
\end{equation}
Here $G=\rm d ln \emph{B}/\emph{R} \rm d \phi$ represents the magnitude of magnetic field gradient. The Alfv\'{e}n and the Hall velocities are defined as $v_{\rm
A}^{2}=B^{2}/4\pi \rho$ and $v_{\rm H}^{2}=\mathbf{\Omega} \cdot \textbf{B}c/2\pi en_{\rm e}$. The disc becomes unstable if any of the roots of the Eq. (11) has a positive real part. We seek the solution at a fiducial radius (\emph{R}). In this investigation, the fiducial radius corresponds to the corotation radius where Keplerian angular velocity in the disc
equals the angular velocity of the star. We first solve the dispersion relation for the limit $\eta \longrightarrow 0$ (zero resistivity). Then, we shall present more general solutions with finite resistivity.

\section{Linear growth rates}
\subsection{Dispersion relation and the growth rates for $\eta = 0$}
\begin{figure}[!ht]
  \begin{center}
    \begin{tabular}{l}
      {\includegraphics[angle=270,scale=0.38]{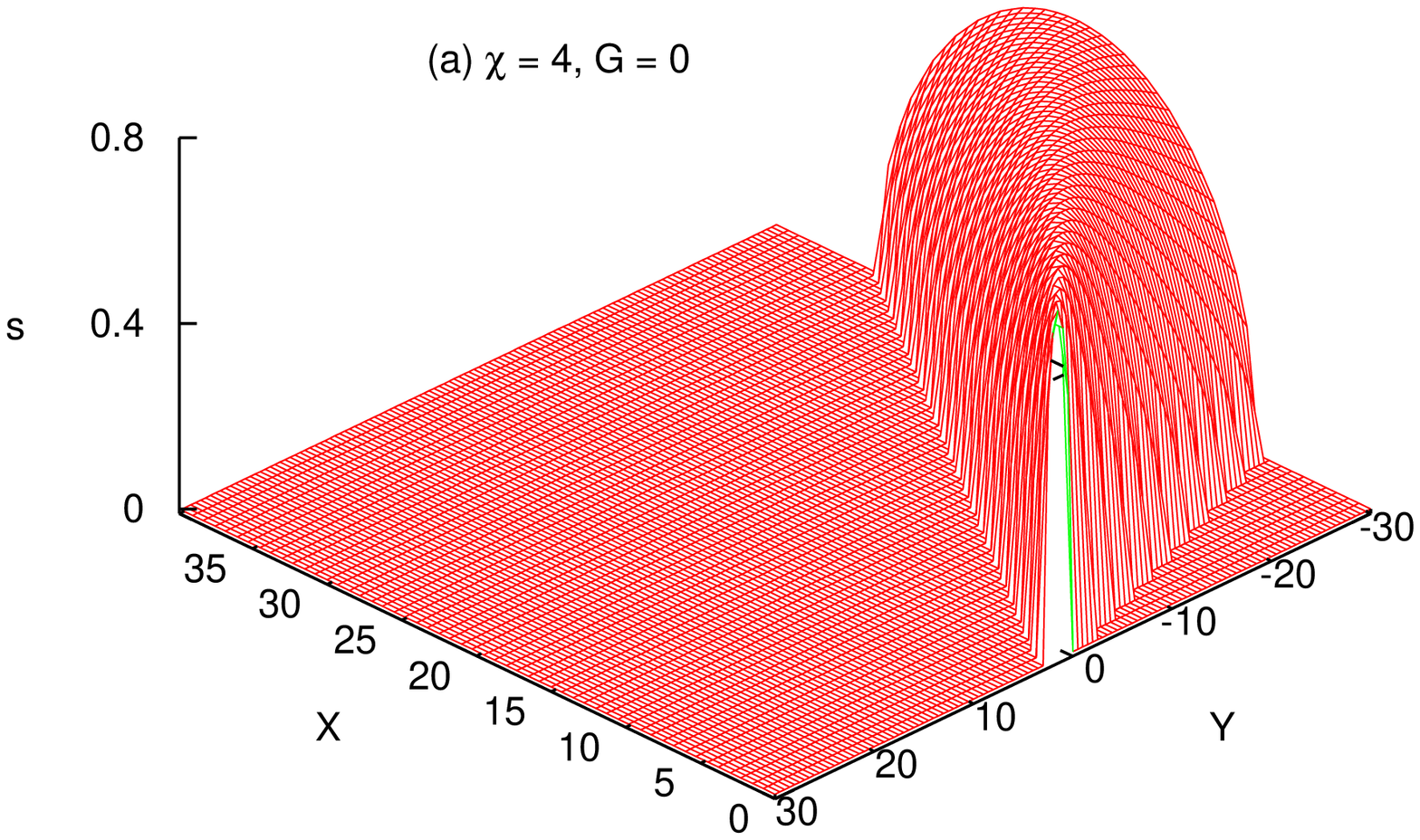}}\\
      {\includegraphics[angle=270,scale=0.38]{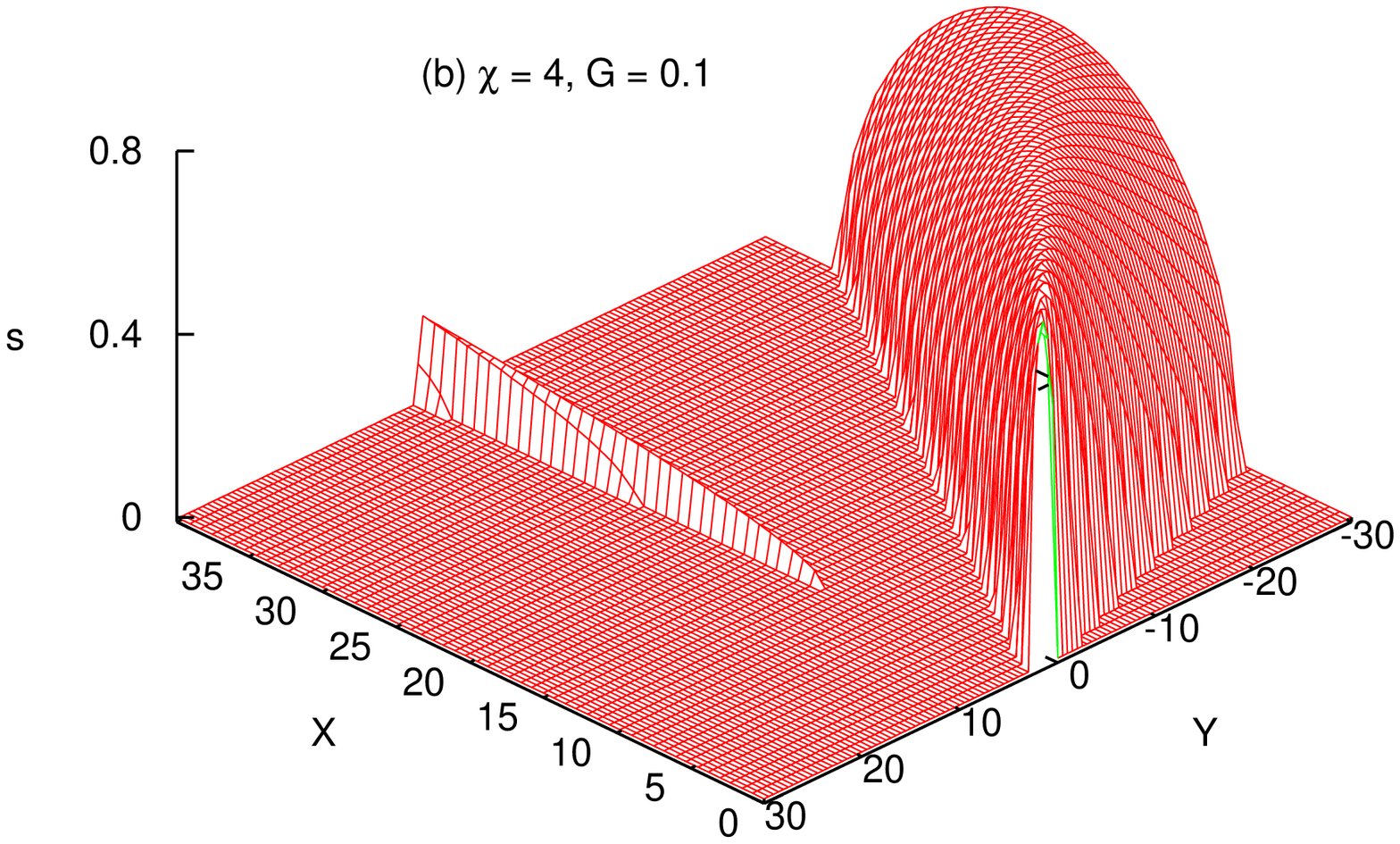}}\\
      {\includegraphics[angle=270,scale=0.38]{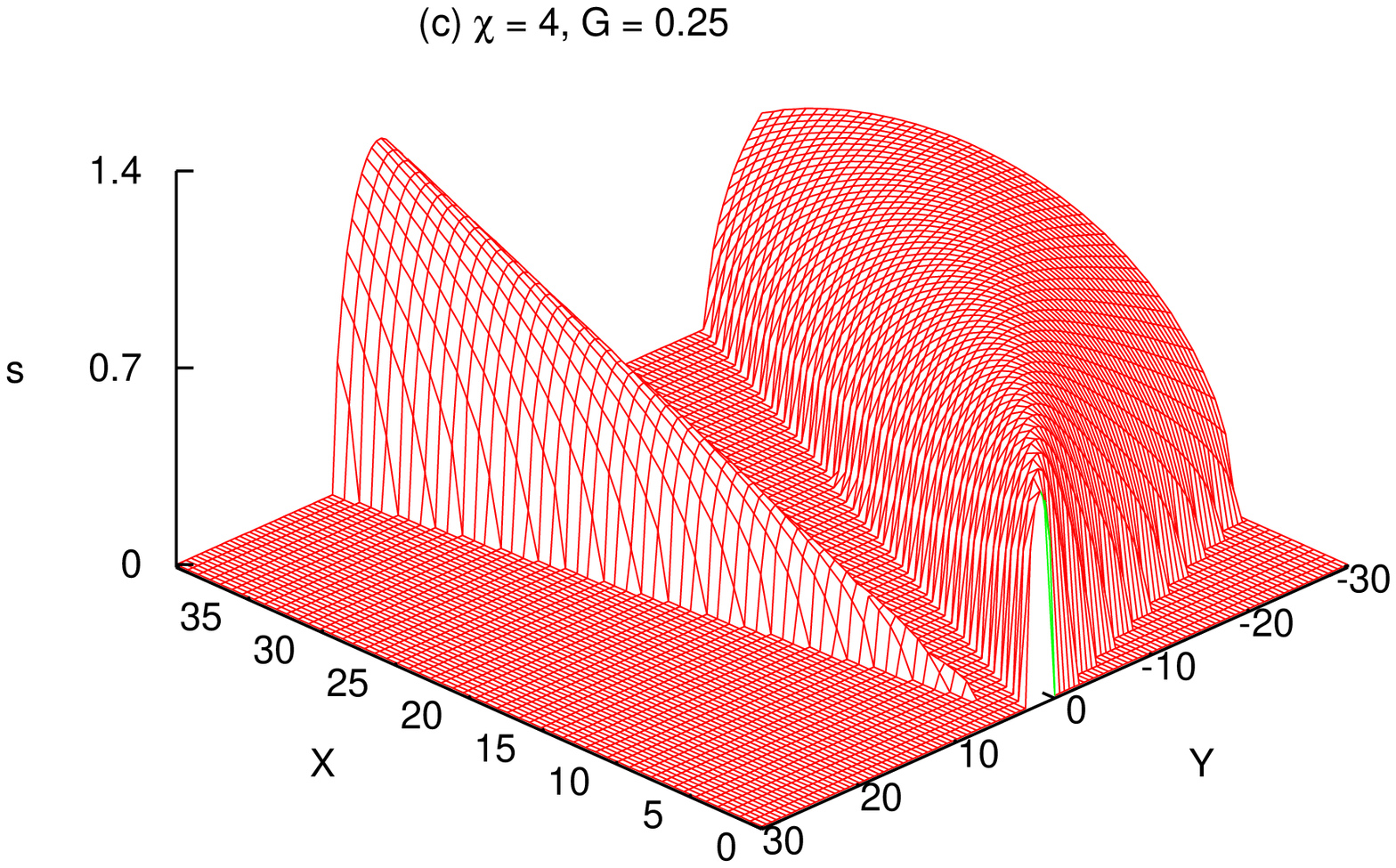}}\\
                \end{tabular}
    \caption{Growth rates found from Eq. (12) for $\mathbf{\Omega}\cdot \mathbf{B} > 0$.
    In order to see the effect of $\nabla B$ on growth rates, we plot our graphs for $\tilde{G}$ = 0, 0.1, 0.25 (see text for definitions). A new unstable mode comes into existence with the inclusion of
$\nabla B$. See Table 1 for maximum values of the growth rates
($s_{\rm m}$).}
     \end{center}
\end{figure}

To write Eq. (11) in dimensionless form, we normalize the growth rate, Alfv\'{e}n and Hall velocities, epicyclic frequency, the magnetic field gradient and the Hall parameter as follows: \emph{s} = $\omega/\Omega$,  X = $(kv_{A}/\Omega)^{2}$, Y =
$(kv_{H}/\Omega)^{2}$,  $\tilde{\kappa} = \kappa/\Omega$, $\tilde{G} = G/k$,
$\chi \equiv v_{H}^{2}/v_{A}^{2}$. Therefore, in the limit of zero resistivity ($\eta = 0$) the dispersion relation (11) can be rewritten in terms of dimensionless parameters as
\begin{equation}
\label{drI}
\begin{array}{l}
{s^4+s^2\left[{\tilde{\kappa }}^2 +  X(2-\tilde{G}^2)+\dfrac{{Y}}{4}\left(\dfrac{d\ln \,\Omega^2}{d\,\ln \,R}+ Y-\chi X\tilde{G}^2 \right)\right]}\\
\\

{+\left( X + \dfrac{{ Y}{\tilde{\kappa }}^2}{4}\right)\left[ \dfrac{ d \,\ln \,\Omega^2}{ d \,\ln \,R}+( X+ Y)(1-\tilde{G}^2)
\right]=0.}
\end{array}
\end{equation}
\\
We note that Eq. (12)
is reduced to Eq. (62) of \cite{BT2001} when $\tilde{G}$ = 0. We can easily calculate $\tilde{G}$ values by using the magnetic dipole formulae which give the magnetic field strength depending on radial distance and the magnetic latitude. Assuming that the equatorial magnetic field strength on stellar surface $B_*$=1 kG and bearing that $kR \gg 1$ in mind, we find the maximum value for $\tilde{G}$ as about 0.25 for moderate inclinations. We assume a Keplerian rotation, therefore the dimensionless epicyclic frequency is $\tilde{\kappa} = 1$. The Hall parameter for T Tauri disks can be estimated as follows (see also DP2012):
The radial distance where the stellar magnetic field starts to control the motion of the accreting plasma is given by \cite{Bouvier2007a} as about 7 stellar radii for $B_*$= 1 kG. Then the magnetic field strength at $R_{\rm M}=7 R_{*} $ is $B_{\rm M}=B_{*}(R_{*}/R_{\rm M})^{3}\approx 2.9$ G. If we take $\rho_g=10^{-9} \rm gcm^{-3}$ \citep{Alexander2008} and $n_e=10^5 \rm cm^{-3}$ \citep{Glassgold2007}, we find the $\chi$ value as 4 for a typical period $P \sim 8^{\rm d}$. We will use these values to solve the dispersion relation given by (12) numerically. The solution of Eq. (12) gives two fast and two slow modes which are labeled according to the magnitude of their phase velocities, $v_{ph} = \omega_i/k$. One of the slow modes is unstable. The graphical solutions of Eq. (12), the numerical growth rates (\emph{s}) of the slow mode, are thus shown in (X,Y) plane in Fig. 1. Positive \emph{s} implies an unstable exponential growth of the mode. The regions of instability are therefore seen
as \textquotedblleft ridges\textquotedblright\, above the (X,Y) plane. We first assume that $\mathbf{\Omega}\cdot \mathbf{B} > 0$. In order to see the effect of the azimuthal component of the magnetic field gradient on growth rates, we plot our graphs for weak ($\tilde{G}$ = 0.1) and strong ($\tilde{G}$ = 0.25) gradient. The standard MRI analyses show one unstable region in the (X,Y) plane when the magnetic field gradients are not included (i.e. $\tilde{G}$ = 0). Fig. 1a shows the unstable region found from previous MRI analyses which are not modified by the field gradient. From now on, we refer to this region of instability as the \emph{unstable region I} (\emph{UR-I}). The maximum growth rate of the instability for this case has been found to be 0.75. Fig. 1b and 1c show that the inclusion of the magnetic field gradient in the azimuthal direction produces a new unstable region in the (X,Y) plane. We refer to this new region of instability as the \emph{unstable region II} (\emph{UR-II}). Fig. 1b shows the growth rates for $\tilde{G}$ = 0.1. Here, the maximum value of the growth rate for \emph{UR-I} is 0.77. If we compare Fig. 1b and Fig. 1a, we see that the region of instability becomes wider with the inclusion of the gradient. The maximum value of the growth rate for \emph{UR-II} is smaller than that of \emph{UR-I}. Fig. 1c shows the growth rates for strong gradient ($\tilde{G}$ = 0.25). It is apparent that the strong gradient produces higher growth rates and larger unstable regions in the (X,Y) plane. Moreover, the growth rates of \emph{UR-II} becomes higher than that of \emph{UR-I} when $\tilde{G}$ = 0.25. The maximum growth rate which is included in \emph{UR-II} reaches a value of 1.72. The maximum growth rates of the \emph{UR-I} and the \emph{UR-II} are listed in Table 1 for different values of $\tilde{G}$. The growth rates of the \emph{UR-II} are affected by the magnitude of the gradient more than that of \emph{UR-I}.

Fig. 2 shows the growth rates for $\mathbf{\Omega}\cdot \mathbf{B} < 0$. \emph{UR-II} again spreads over a larger space in the (X,Y) plane for $\tilde{G}$ = 0.25 than it does for $\tilde{G}$ = 0.1. But, \emph{UR-I} becomes smaller when the gradient is strong. When we compare Fig. 1 and Fig. 2, we see that the maximum values of the growth rates are very similar. However, the wavenumber range where the instability occurs is narrower in Fig. 2. The difference in unstable regions between two figures is much more apparent in \emph{UR-I}.

\begin{figure}[!ht]
  \begin{center}
    \begin{tabular}{l}
      {\includegraphics[angle=270,scale=0.38]{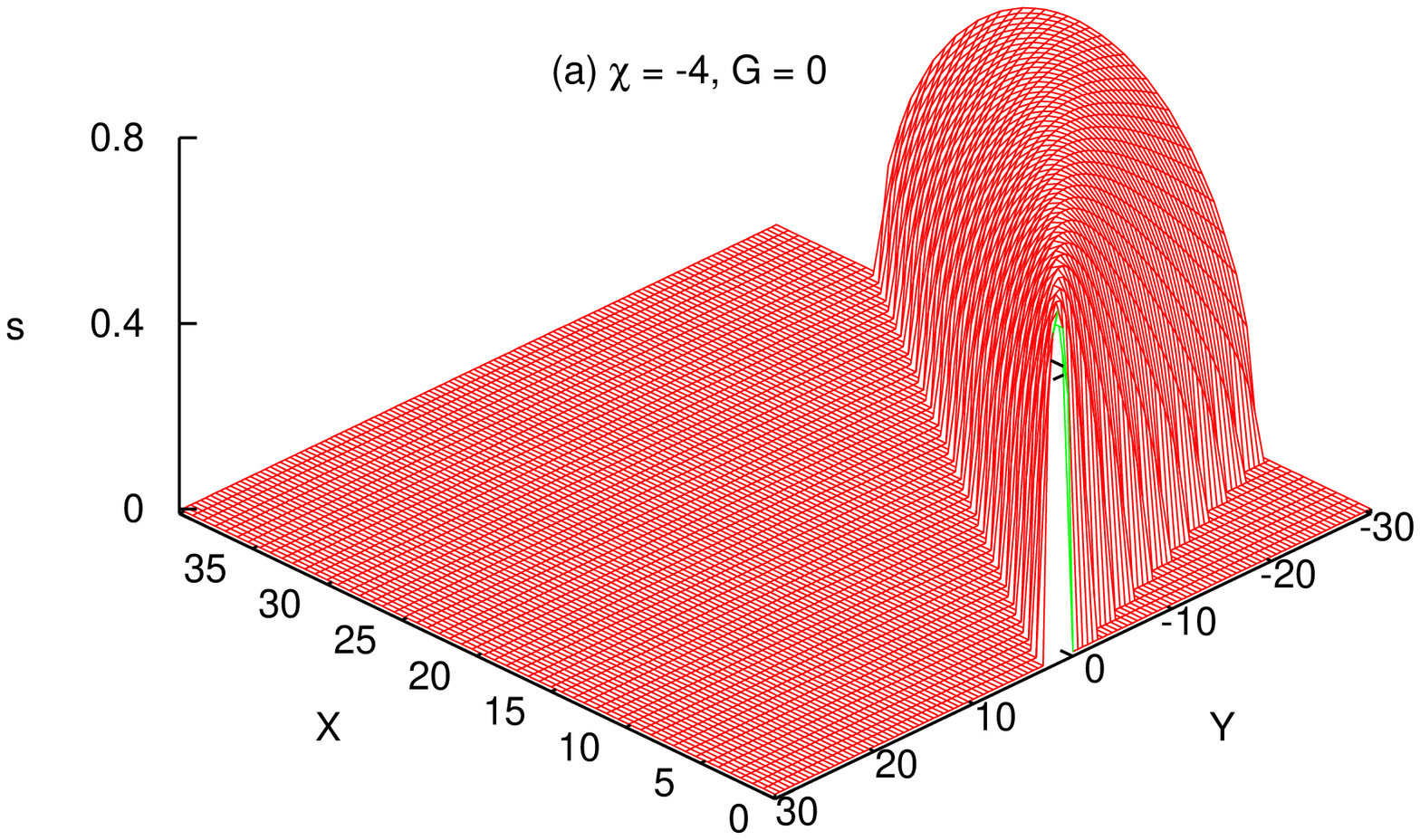}}\\
      {\includegraphics[angle=270,scale=0.38]{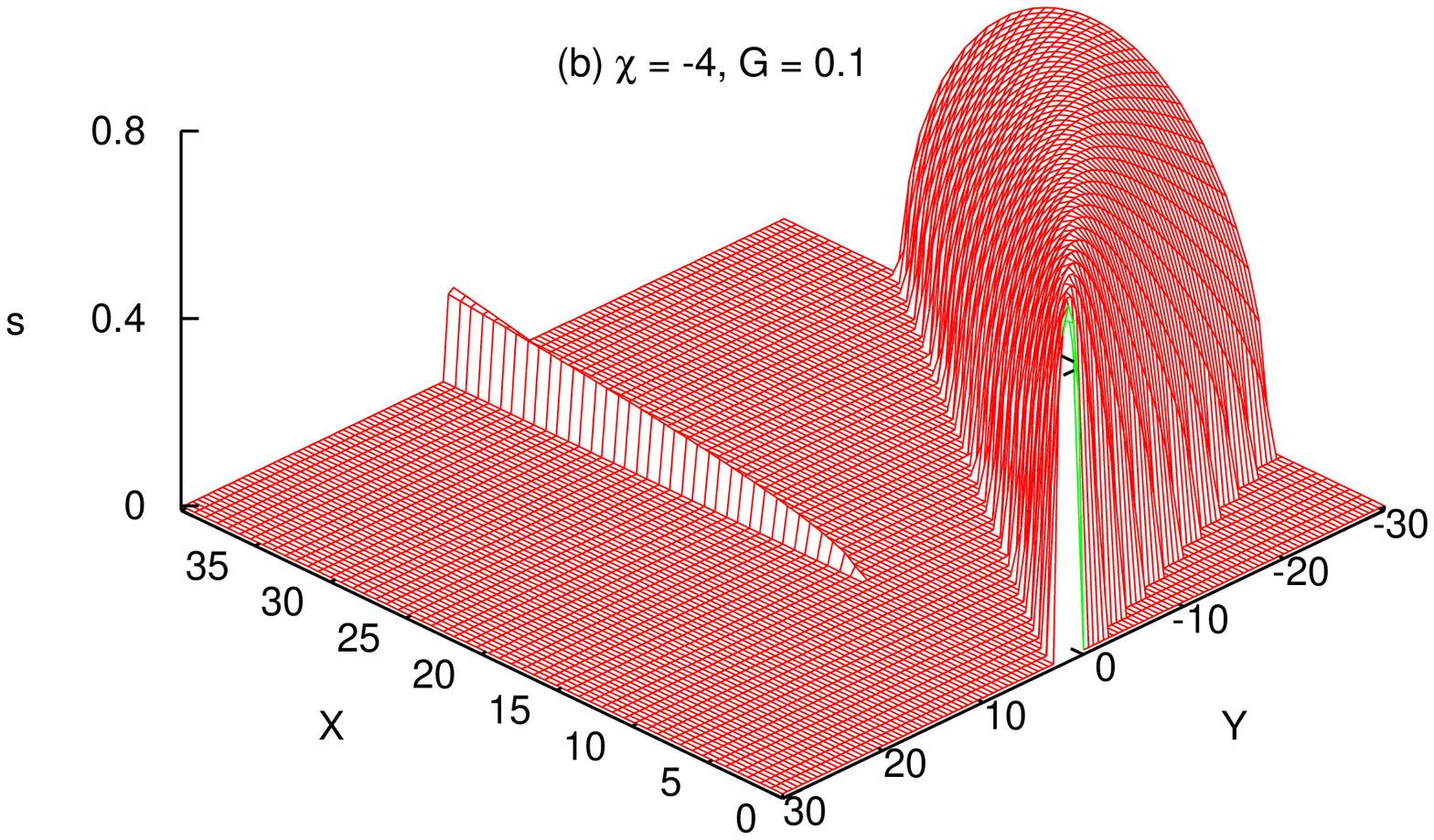}}\\
      {\includegraphics[angle=270,scale=0.38]{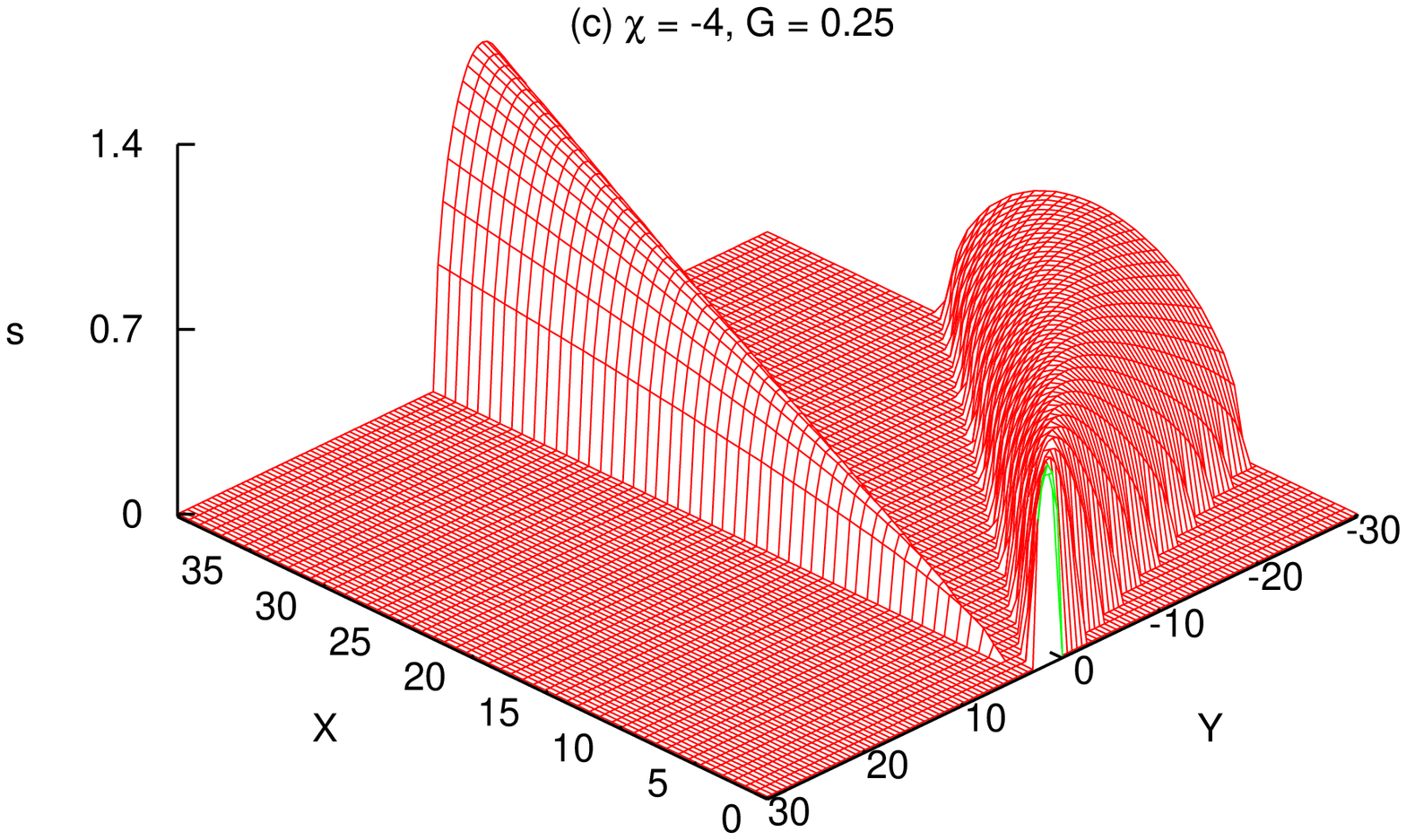}}\\
                \end{tabular}
    \caption{Growth rates found from Eq. (12) for $\mathbf{\Omega}\cdot \mathbf{B} < 0$.
    In order to see the effect of $\nabla B$ on growth rates, we plot our graphs for $\tilde{G}$ = 0, 0.1, 0.25 (see text for definitions). A new unstable mode comes into existence with the inclusion of
$\nabla B$. See Table 1 for maximum values of the growth rates
($s_{\rm m}$).}
     \end{center}
\end{figure}

\begin{table}
\caption{The maximum growth rates of the instability.} % title of Table
\begin{center}
\begin{tabular}{lcccc}
\hline
%&   &    &    &   &   & \\
& \multicolumn{4}{c}{$s_{\rm m}$}\\
\hline
%\cline{1-10}
& \multicolumn{2}{c}{$\chi = 4$} & \multicolumn{2}{c}{$\chi = -4$}\\
 & UR-I & UR-II & UR-I & UR-II \\
%&   &    &   &   &   &   &   &   & \\
\hline % inserts single horizontal line
   &   &   &   & \\
$\tilde{G}$ = 0&0.75& - &0.75& - \\
$\tilde{G}$ = 0.1&0.77&0.19&0.75&0.19\\
$\tilde{G}$ = 0.25&0.89&1.19&0.75&1.24\\
&   &   &   & \\
\hline
\end{tabular}
%\medskip
\end{center}
\end{table}

\subsection{Dispersion relation and the growth rates for $\eta \neq 0$}

In the presence of resistivity the dispersion relation can be written in the dimensionless form as
\begin{equation}
\begin{array}{l}
{s^5+\dfrac{ 3X}{ Re_m}s^4+s^3\left\{\begin{array}{l}{{\tilde{\kappa}}^2+X\left(\dfrac{ 3X}{ Re_m}+2-\tilde{G}^2\right)}\\
\\
{+\dfrac{ \chi X}{ 4} \left(\dfrac{  d \ln \,\Omega^2}{ d \ln R} + (\chi-\tilde{G}^2)X \right)}\end{array}\right\}}\\
\\
{+s^2\dfrac{ X}{ Re_m}\left\{\begin{array}{l}{3{\tilde{\kappa}}^2+X\left(\dfrac{X}{Re_m}+4-2\tilde{G}^2\right)}\\
\\
{+\dfrac{ \chi X}{ 4} \left(\dfrac{ d \ln \,\Omega^2}{ d \,\ln\, R} + (\chi-\tilde{G}^2)X \right)}\end{array}\right\}}\\
\\
{+s\left\{\begin{array}{l}{\left(X+\dfrac{ Y\kappa^2}{ 4} \right)\bigg[\dfrac{ d \ln \,\Omega^2}{ d \ln \,R}+X\bigg( (\chi+1)(1-\tilde{G}^2)\bigg)\bigg]}\\
\\
{+\dfrac{ X^2}{ Re_m^2}\left(\tilde{\kappa}^2+X(2-\tilde{G}^2)\right)}\end{array}\right\}}\\
\\
{+X\left(\dfrac{ 1}{ Re_m}+\dfrac{ \kappa^2 \chi }{ 4Re_m}\right)\bigg[\dfrac{ d \ln \,\Omega^2}{ d \ln R}+X\bigg( (\chi+1)(1-\tilde{G}^2)\bigg)\bigg]=0.}
\end{array}
\end{equation}

\begin{figure*}[!ht]
  \begin{center}
    \begin{tabular}{ll}
      {\includegraphics[angle=270,scale=.4]{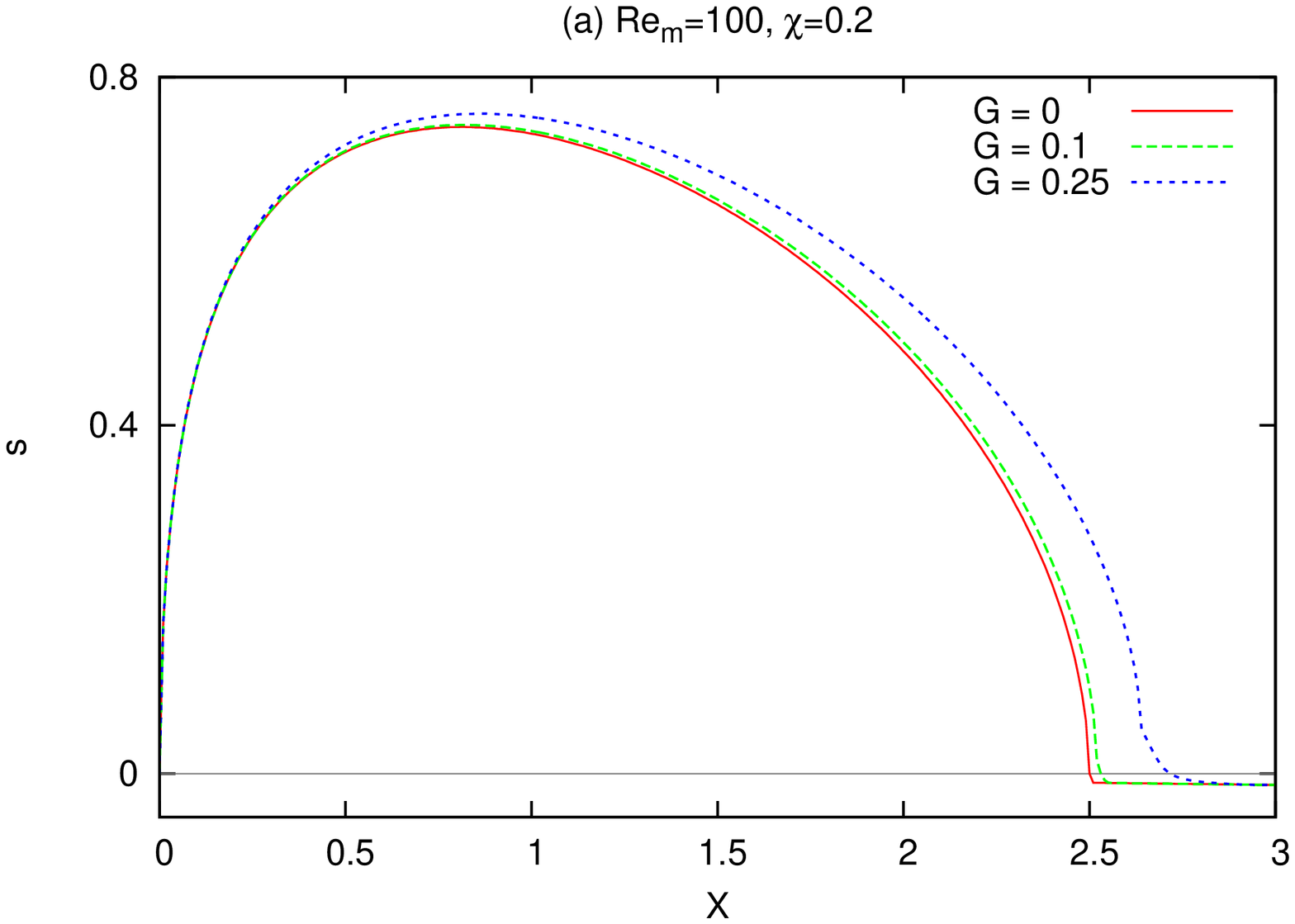}}&
      {\includegraphics[angle=270,scale=.4]{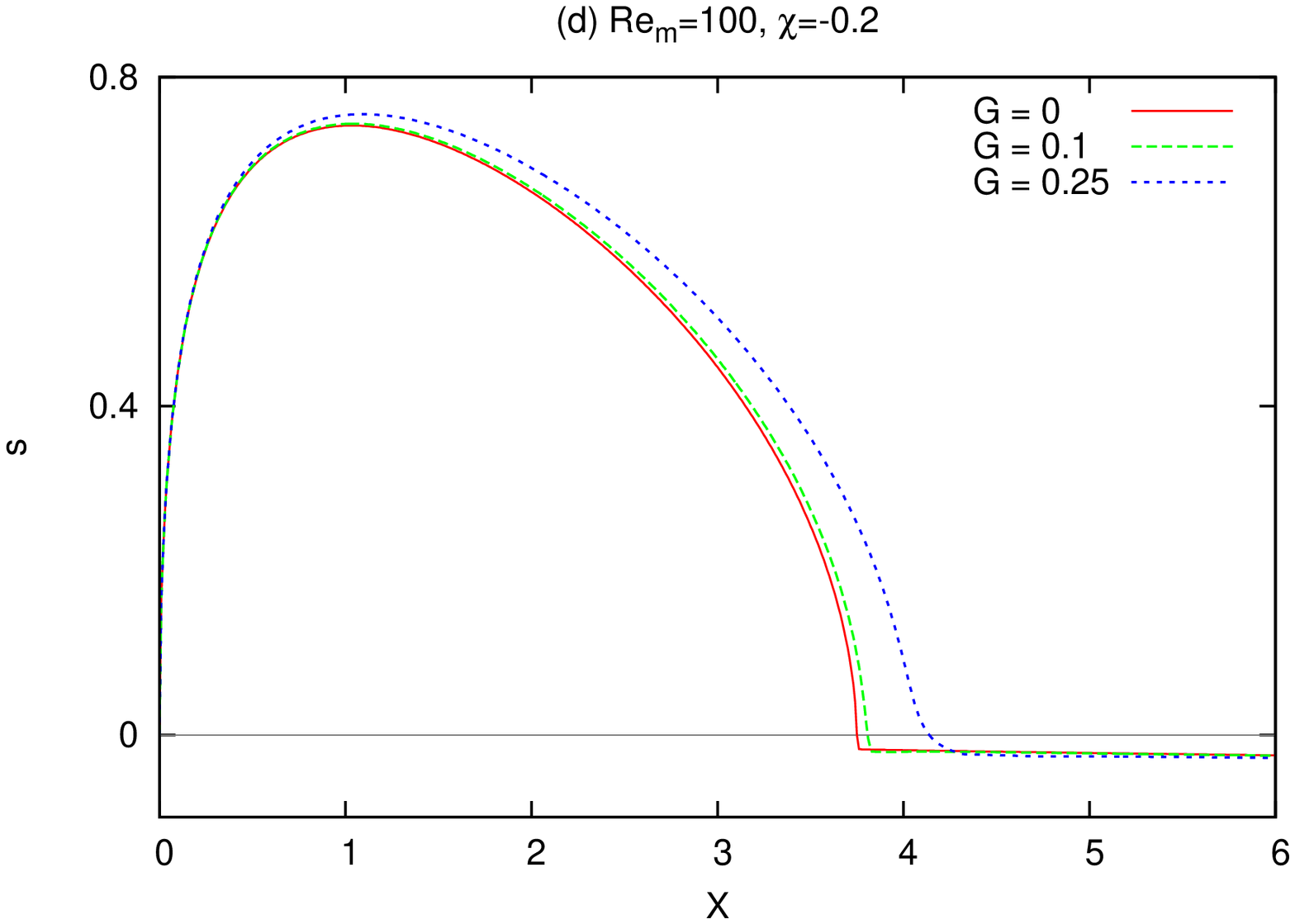}}\\
      {\includegraphics[angle=270,scale=.4]{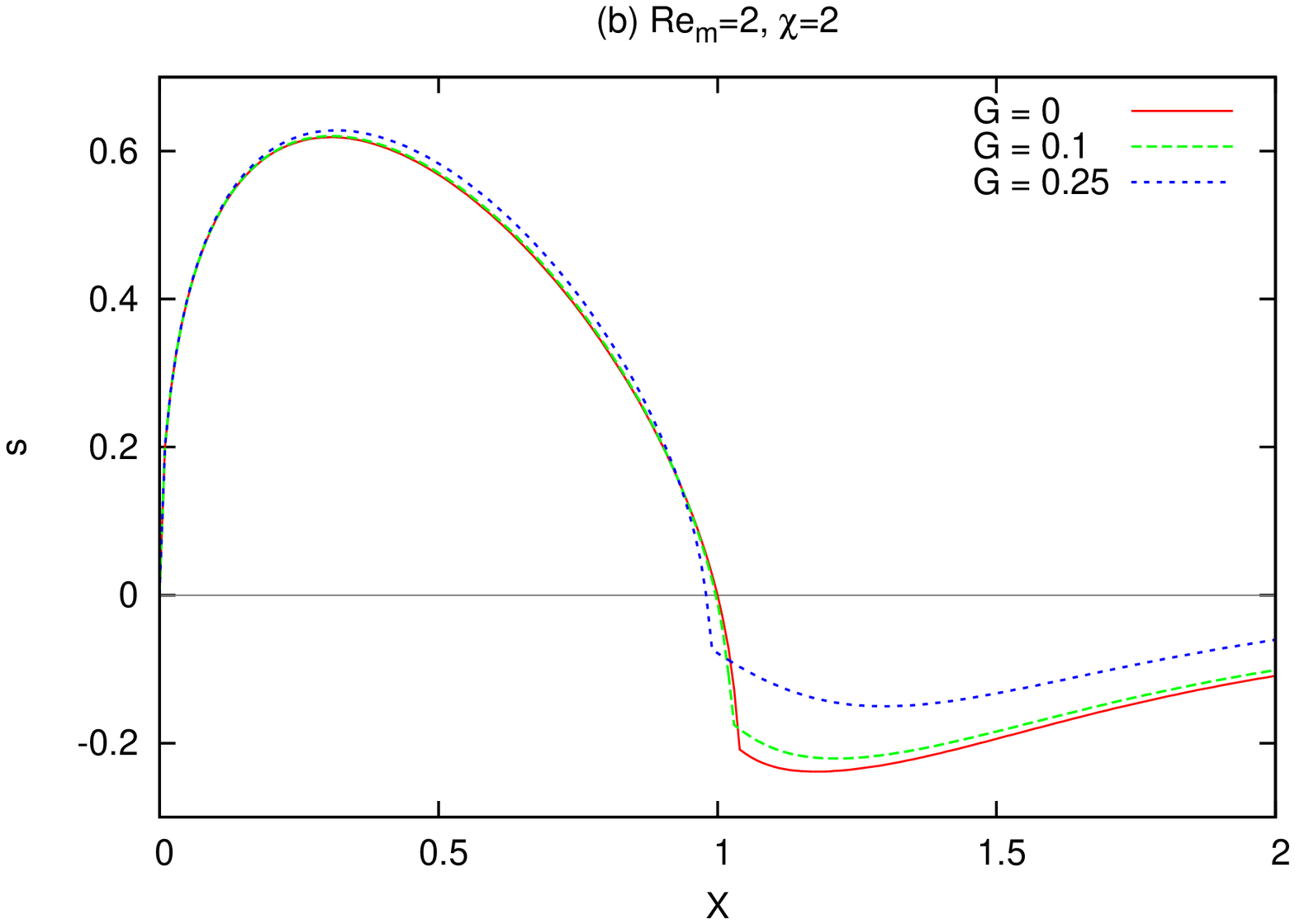}}&
      {\includegraphics[angle=270,scale=.4]{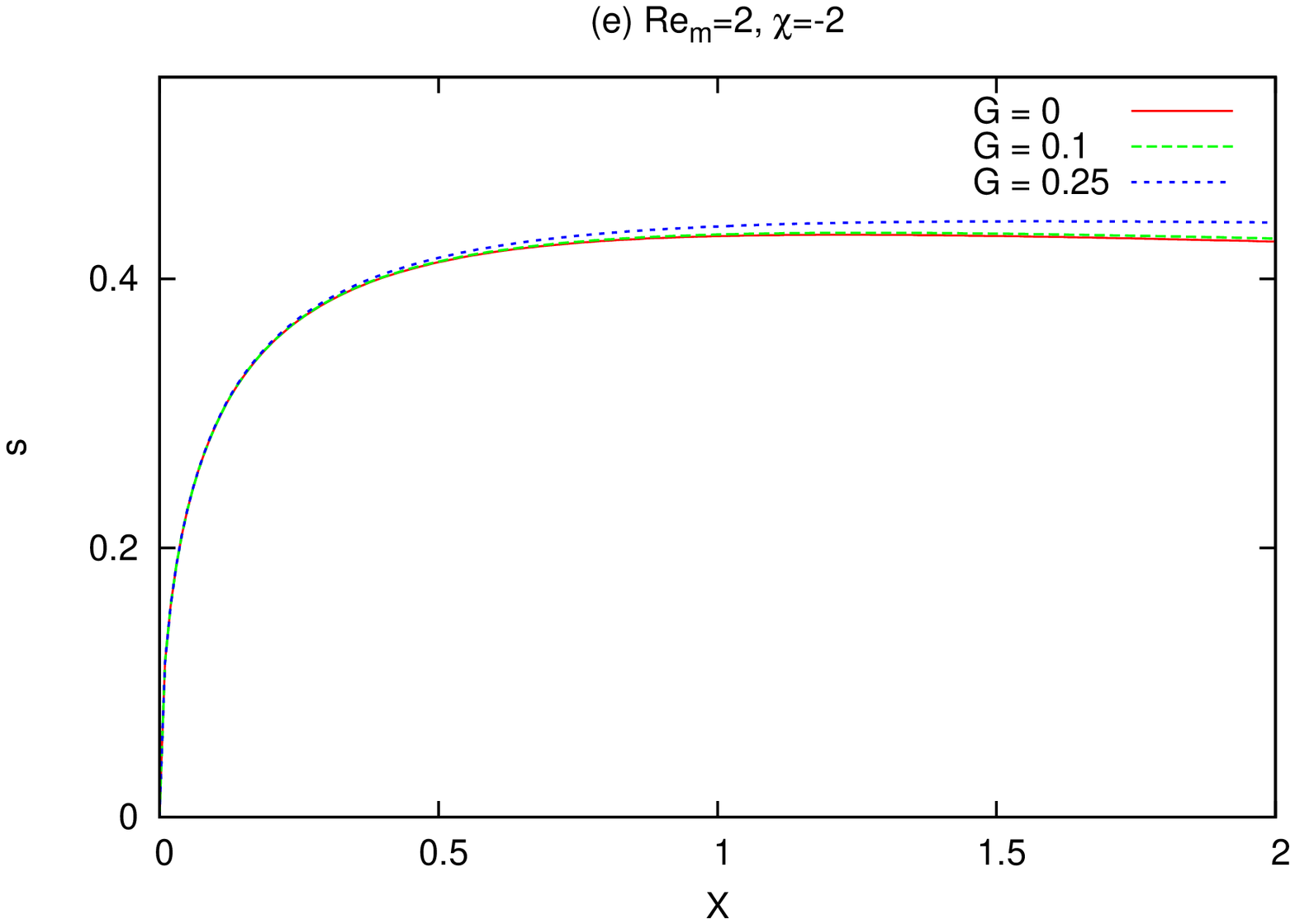}}\\
      {\includegraphics[angle=270,scale=.4]{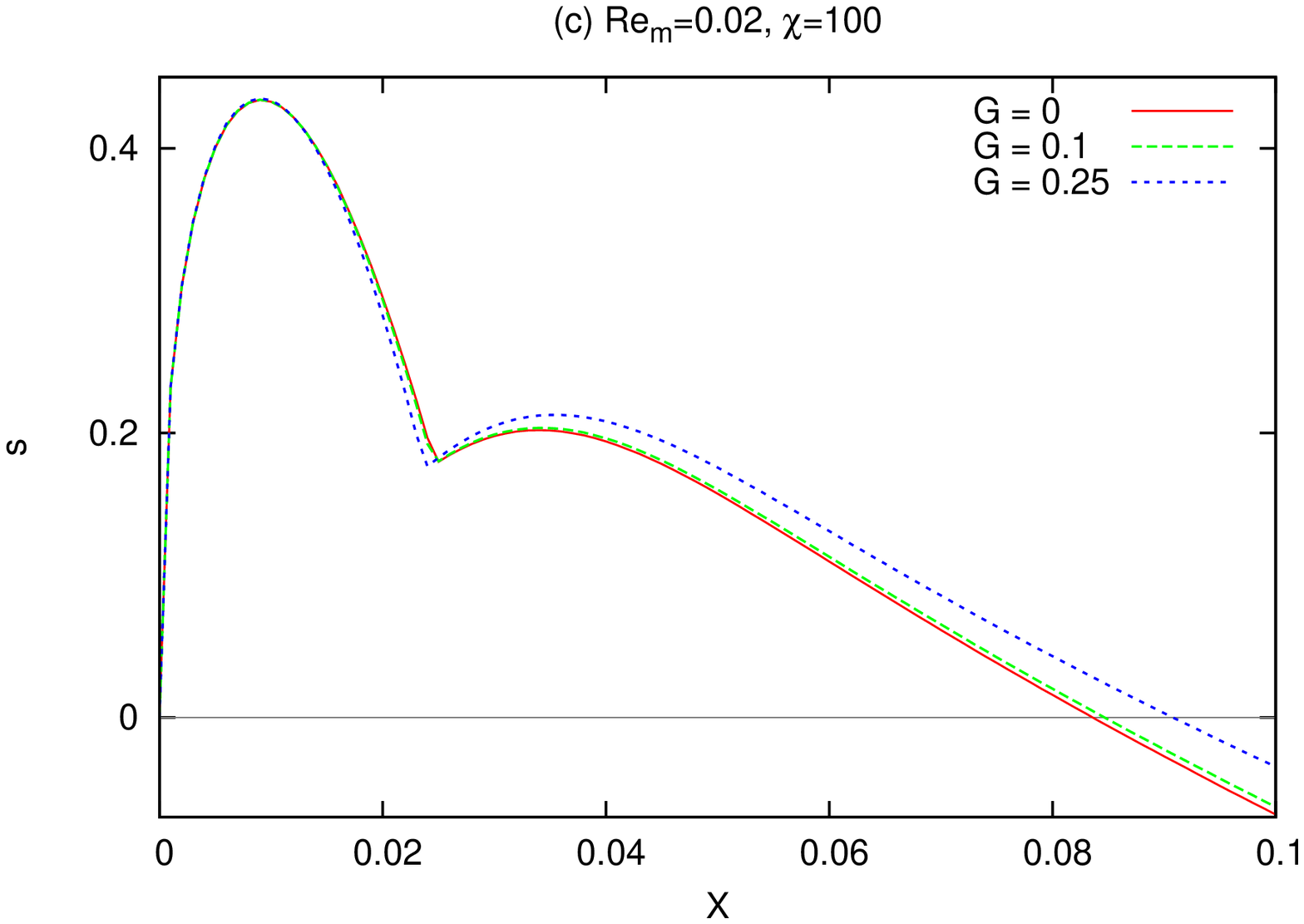}}&
      {\includegraphics[angle=270,scale=.4]{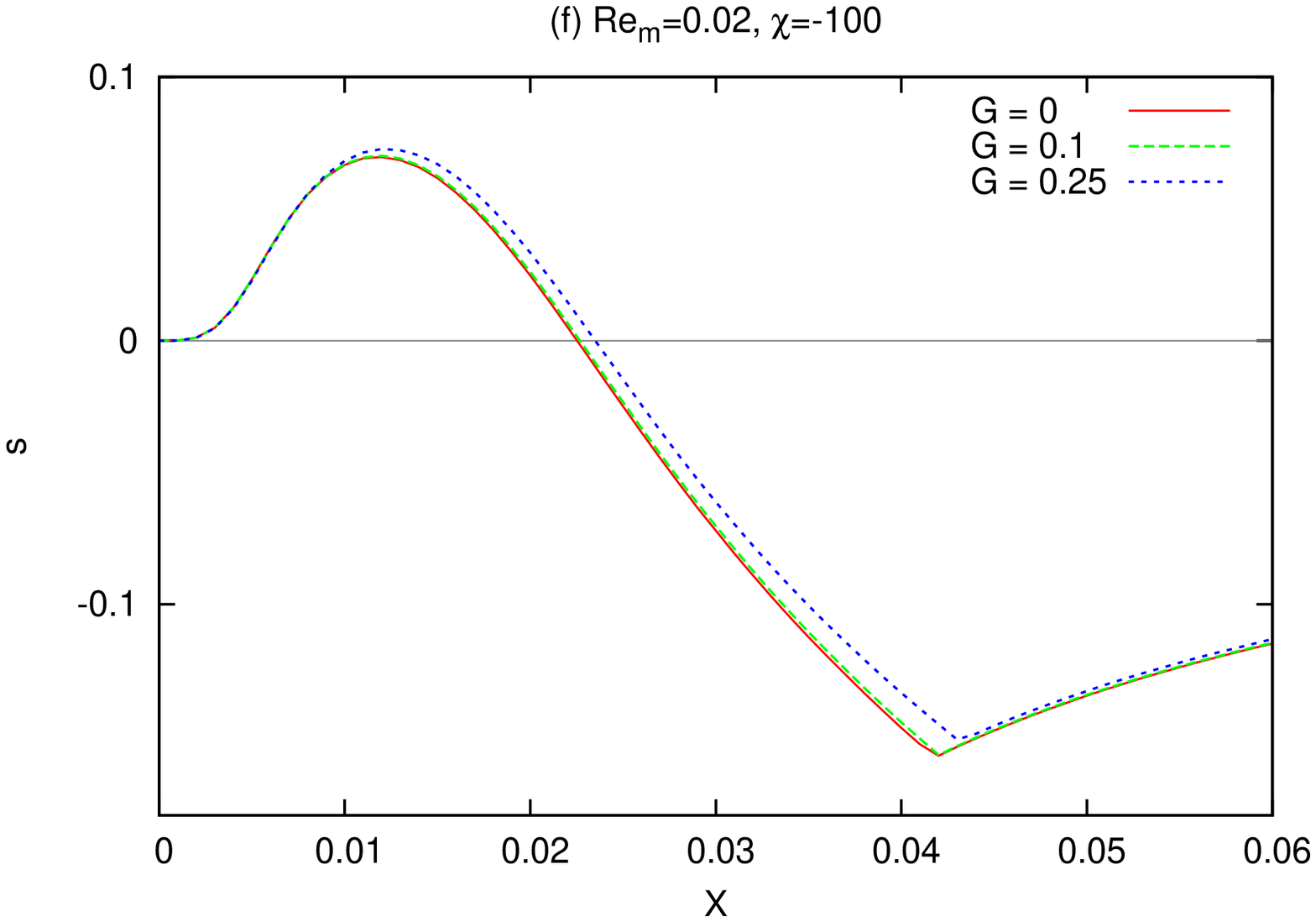}}\\

    \end{tabular}
    \caption{Growth rates found from Eq. 13.
    The figures are plotted for $\tilde{G} =$ 0, 0.1, 0.25.
    Left panels are for $\mathbf{\Omega}\cdot \mathbf{B} > 0$;
the right panels are for $\mathbf{\Omega}\cdot \mathbf{B} < 0$.}
     \end{center}
\end{figure*}

Here, $Re_M \equiv v_A^2/ \Omega\eta $ is the magnetic Reynolds number. In order to solve Eq. (13) numerically, we need to make assumptions on the relative importance of the non-ideal MHD terms. \cite{SS2002} show that the PPDs can be separated into three regions which are classified depending on the relative importance of the Hall effect: (i) the outer region of the disc where $|\chi| < 1$ and $Re_M \gg 1$ (i.e. both the Hall effect and ohmic dissipation are unimportant), (ii) the intermediate region of the disc where $|\chi| > 1$ and $Re_M > 1$ (i.e. the Hall effect is important but ohmic dissipation is still can be neglected), (iii) the inner part of the disc where $|\chi| \gg 1$ and $Re_M \ll 1$ (i.e. both effects are important). We investigate the effect of the magnetic field gradient on MRI in three regions of the disc.

Eq. (13) gives two fast, two slow and one standing wave modes. One of the slow modes is unstable. The right panel of Fig. 3 shows the dispersion relations for $\mathbf{\Omega}\cdot \mathbf{B} > 0$. To see the effect of the magnetic field gradient, we plot our graphs for $\tilde{G} = 0, 0.1, 0.25$. Modes with smaller wavenumber than some critical value ($k_c$) are unstable, modes exceeding the critical value are stable for the MRI. Fig. 3a shows that, in the outer region where we assume that $\chi = 0.2$, $Re_M = 100$, the magnetic field gradient extends the wavenumber range of the instability to higher values and also produces higher growth rates. Fig. 3b shows the growth rates for $\chi = 2$, $Re_M = 2$ and $\tilde{G} = 0, 0.1, 0.25$. Here, the growth rates are reduced by the effect of ohmic dissipation. The critical wavenumbers are also found to be smaller than the first case. But the larger magnetic field gradient again enhances the growth rates and the critical wavenumber. As shown in Fig. 3c, decreasing $Re_M$ leads to even smaller growth rates and critical wavenumbers. Here, the growth rates are plotted for $\chi = 100$, $Re_M = 0.02$ and $\tilde{G} = 0, 0.1, 0.25$. In this case, the magnetic field gradient has very little effect on the maximum growth rate, but increasing the field gradient enhances the critical wavenumber of the instability.

The left panel shows the dispersion relations for $\mathbf{\Omega}\cdot \mathbf{B} < 0$. If we compare Fig. 3a and 3d, we see that negative $\chi$ produces wider unstable regions. Fig. 3e and 3f show that decreasing the $Re_M$ (i.e. increasing the effect of ohmic dissipation) reduces the growth rates significantly. There is no characteristic scale for the instability as the critical wavenumber goes to infinity when $\chi = -2$, $Re_M = 2$.

In conclusion, the net effect of the magnetic field gradient is to shift the critical wavenumber towards higher values. This is an expected result, as the role of the magnetic field strength is to determine a characteristic scale for unstable wavelengths, a scale that makes \emph{k} comparable to $\Omega/v_A$ (\citealt{BH1991,BH1998}). If the magnetic field is weakened, then the critical wavenumber required for the MRI increases. In Fig. 3, the curves with $\tilde{G} \neq 0$ show this effect. On the other hand, increasing the non-ideal MHD effects produces much smaller growth rates and critical wavenumbers. Fields oriented such that $\mathbf{\Omega}\cdot \mathbf{B} < 0$ extends the wavenumber range of the unstable region when the ohmic dissipation is not important. But the growth rates are always higher when $\mathbf{\Omega}\cdot \mathbf{B} > 0$.

\section{Discussion}
\label{results}
Any realistic model on accretion onto a central object with a dipole magnetic field should take into account the effects of misalignment of the magnetic axis. Previously, it has been shown that the inclusion of the magnetic field gradient in the radial direction plays an important role in triggering MRI. We now show that the magnetic field gradient in the azimuthal direction also makes MRI more powerful and produces a new unstable region. Inclusion of the magnetic field gradient influences the growth rates and the characteristic wavenumbers significantly. While increasing the non-ideal MHD effects suppresses the growth of the instability, increasing the magnitude of the magnetic field gradient produces higher growth rates. The behaviour we have discussed in this paper is relevant to T Tauri discs where the stellar magnetic axis is misaligned with respect to the rotation axis. For example, AA Tau is known to have a 2 - 3 kG magnetic dipole tilted at about $20^{\circ}$ to the rotation axis \citep{Bouvier2007b}. If we naively assume that the rotation axis of the star and the disc coincide, we find the variation in the field strength in the azimuthal direction that the disc particles experience during their orbital motion as 14 per cent for this inclination. This variation gives $\tilde{G} \approx 0.1$ which corresponds to the case we considered in our numerical results.

Previously, \cite{DP2007} and DP12 showed that MRI becomes more powerful with the inclusion of the magnetic field gradient produced by magnetization currents in the radial direction. If we compare the effects of radial and azimuthal gradients on MRI, we can see that both gradients produce a new unstable region. The maximum values of the growth rate found from these new unstable regions are very similar for a weak gradient in the radial and azimuthal direction. The maximum growth rate is 0.19 for $\tilde{G}=0.1$ and 0.22 for $G = 0.1$ where \emph{G} is the radial gradient in dimensionless form, i.e. $G=d ln B/d ln R$ (cf. DP12, Table 1). We should note that the magnitude of the azimuthal gradient caused by the misalignment of dipole field may not be as high as the radial gradient  produced by magnetization. However, the influence of a strong azimuthal gradient on the new unstable region seems to be slightly stronger than that of a radial gradient. Here, we find a wider unstable region and a higher maximum growth rate from UR-II for $\tilde{G}=0.25$ than we found for $G = 0.5$ (cf. DP12, Fig. 4b). In order to make a comparison, we solved Eq. 18 of DP12 for $G = 0.25$ and found the maximum growth rate as 0.94 which is slightly lower than we found in this study for $\tilde{G}=0.25$ (see Table 1). It is a condition for the growth of MRI that the field be weak and both investigations reveal that weakening an already weak magnetic field triggers MRI and enhances the growth rate of the instability. In conclusion, the inclusion of magnetic field gradient tends to make the instability more powerful.

Previously, \cite{BT2001} pointed out the importance of identifying the Hall parameter precisely. When the Hall term is defined depending on the product $\mathbf{\Omega}\cdot \mathbf{B}$, this raises the following question: what happens when the magnetic field lies in the disc plane, i.e. $\mathbf{\Omega}\cdot \mathbf{B} = 0$? To answer this question, BT2001 defined the Hall parameter more precisely and showed that the coupling is actually $(\textbf{k}\cdot \mathbf{\Omega})(\textbf{k}\cdot \textbf{B})$. If the magnetic field has a radial component, then it is always possible to find a wavevector that makes the Hall parameter negative. The analysis described here is restricted to axial fields and axial wavenumbers, therefore we could not address this problem. In this work, we consider the vertical component of the magnetic field ($B_z$) only, as it is crucial for MRI. However, if the central object has a misaligned dipole field, then the magnetic field threading the disc will always have a radial component. We shall make more general analysis including the radial field component and explore this problem in a future publication.

In realistic case, the problem is much more complicated than described here. Because the magnetic field may not be purely dipole. The strength and the structure of the magnetic field is very important as it determines the magnetopause radius where the disc plasma is separated from the stellar magnetosphere, and therefore the inner disc radius \citep{Aly1980}. But if the magnetic axis is misaligned with the rotation axis of the disc, we always expect a magnetic field gradient to occur in the disc plane whatever the field structure is. We should also note that a non-axisymmetric field is expected to generate non-axisymmetric perturbations. In this paper, we restricted our attention to a magnetic field subject to axisymmetric perturbations. Previously, it has been demonstrated that non-axisymmetric MRI modes can appear as transient events only \citep{BH1992,TP1996}. However, exponentially growing non-axisymmetric MRI modes have also been obtained by \cite{KR2010} for Keplerian discs. For non-axisymmetric field configurations, it may be argued that non-axisymmetric modes may be more important than the axisymmetric modes, at least transiently. A full stability analysis of a disc in a misaligned dipole field would require the derivation of a dispersion relation for a general perturbation and this will be the purpose of our future work.

\acknowledgements
I would like to thank the referee for his/her valuable suggestions which helped to improve the manuscript, and Prof. Rennan Pek\"{u}nl\"{u} for his useful comments. I thank the Theoretical Astrophysics Group at University of Leicester for their hospitality. This work was supported by The Scientific and Technological Research Council of Turkey (T\"{U}B\.{I}TAK) through the Postdoctoral Research Fellowship Programme (2219).

% Example of using BiBTeX (plus natbib):
% For details see \cite{1999MNRAS.309..731B},
% \cite{1893PASP....5..204C},
% \cite{2008IAUS..252...75L}. It has been demonstrated that this
% is important \citep{2012AN....333..663S}.

% Use this code if you wish to generate your bibliography with BibTeX;
% please replace first the string "an-demo" below with the name(s) of
% the BibTeX data base(s) you want to use.
% The resulting bibliography-output (the contents of the .bbl file)
% must be pasted into this file before submission.
%
\bibliographystyle{an}
\bibliography{bibtex}

\begin{thebibliography}{31}
\expandafter\ifx\csname natexlab\endcsname\relax\def\natexlab#1{#1}\fi

\bibitem[{{Alencar} \& {Batalha}(2002)}]{Alencar2002}
{Alencar}, S.~H.~P. \& {Batalha}, C. 2002, \apj, 571, 378

\bibitem[{{Alexander}(2008)}]{Alexander2008}
{Alexander}, R. 2008, \nar, 52, 60

\bibitem[{{Aly}(1980)}]{Aly1980}
{Aly}, J.~J. 1980, \aap, 86, 192

\bibitem[{{Bai}(2014)}]{Bai2014}
{Bai}, X.-N. 2014, \apj, 791, 137

\bibitem[{{Bai}(2015)}]{Bai2015}
{Bai}, X.-N. 2015, \apj, 798, 84

\bibitem[{{Balbus} \& {Hawley}(1991)}]{BH1991}
{Balbus}, S.~A. \& {Hawley}, J.~F. 1991, \apj, 376, 214

\bibitem[{{Balbus} \& {Hawley}(1992)}]{BH1992}
{Balbus}, S.~A. \& {Hawley}, J.~F. 1992, \apj, 400, 610

\bibitem[{{Balbus} \& {Hawley}(1998)}]{BH1998}
{Balbus}, S.~A. \& {Hawley}, J.~F. 1998, Reviews of Modern Physics, 70, 1

\bibitem[{{Balbus} \& {Terquem}(2001)}]{BT2001}
{Balbus}, S.~A. \& {Terquem}, C. 2001, \apj, 552, 235

\bibitem[{{Bouvier} {et~al.}(2007{\natexlab{a}}){Bouvier}, {Alencar},
  {Boutelier}, {Dougados}, {Balog}, {Grankin}, {Hodgkin}, {Ibrahimov}, {Kun},
  {Magakian}, \& {Pinte}}]{Bouvier2007b}
{Bouvier}, J., {Alencar}, S.~H.~P., {Boutelier}, T., {et~al.}
  2007{\natexlab{a}}, \aap, 463, 1017

\bibitem[{{Bouvier} {et~al.}(2007{\natexlab{b}}){Bouvier}, {Alencar},
  {Harries}, {Johns-Krull}, \& {Romanova}}]{Bouvier2007a}
{Bouvier}, J., {Alencar}, S.~H.~P., {Harries}, T.~J., {Johns-Krull}, C.~M., \&
  {Romanova}, M.~M. 2007{\natexlab{b}}, Protostars and Planets V, 479

\bibitem[{{Devlen} \& {Pek{\"u}nl{\"u}}(2007)}]{DP2007}
{Devlen}, E. \& {Pek{\"u}nl{\"u}}, E.~R. 2007, \mnras, 377, 1245

\bibitem[{{Donati} {et~al.}(2008){Donati}, {Jardine}, {Gregory}, {Petit},
  {Paletou}, {Bouvier}, {Dougados}, {M{\'e}nard}, {Collier Cameron}, {Harries},
  {Hussain}, {Unruh}, {Morin}, {Marsden}, {Manset}, {Auri{\`e}re}, {Catala}, \&
  {Alecian}}]{Donatietal2008}
{Donati}, J.-F., {Jardine}, M.~M., {Gregory}, S.~G., {et~al.} 2008, \mnras,
  386, 1234

\bibitem[{{Do{\v g}an} \& {Pek{\"u}nl{\"u}}(2012)}]{DP2012}
{Do{\v g}an}, S. \& {Pek{\"u}nl{\"u}}, E.~R. 2012, \pasp, 124, 922

\bibitem[{{Gammie}(1996)}]{Gammie1996}
{Gammie}, C.~F. 1996, \apj, 457, 355

\bibitem[{{Gammie} \& {Menou}(1998)}]{GM1998}
{Gammie}, C.~F. \& {Menou}, K. 1998, \apjl, 492, L75

\bibitem[{{Glassgold} {et~al.}(2007){Glassgold}, {Najita}, \&
  {Igea}}]{Glassgold2007}
{Glassgold}, A.~E., {Najita}, J.~R., \& {Igea}, J. 2007, \apj, 656, 515

\bibitem[{{Johns-Krull}(2007)}]{JK2007}
{Johns-Krull}, C.~M. 2007, \apj, 664, 975

\bibitem[{{Johns-Krull} {et~al.}(1999){Johns-Krull}, {Valenti}, {Hatzes}, \&
  {Kanaan}}]{JKetal1999}
{Johns-Krull}, C.~M., {Valenti}, J.~A., {Hatzes}, A.~P., \& {Kanaan}, A. 1999,
  \apjl, 510, L41

\bibitem[{{Kitchatinov} \& {R{\"u}diger}(2010)}]{KR2010}
{Kitchatinov}, L.~L. \& {R{\"u}diger}, G. 2010, \aap, 513, L1

\bibitem[{{Kondi{\'c}} {et~al.}(2012){Kondi{\'c}}, {R{\"u}diger}, {Arlt},
  {R{\"u}diger}, \& {Arlt}}]{Kondicetal2012}
{Kondi{\'c}}, T., {R{\"u}diger}, G., {Arlt}, R., {R{\"u}diger}, G., \& {Arlt},
  R. 2012, Astronomische Nachrichten, 333, 202

\bibitem[{{Kondi{\'c}} {et~al.}(2011){Kondi{\'c}}, {R{\"u}diger}, \&
  {Hollerbach}}]{Kondicetal2011}
{Kondi{\'c}}, T., {R{\"u}diger}, G., \& {Hollerbach}, R. 2011, \aap, 535, L2

\bibitem[{{Kunz} \& {Lesur}(2013)}]{KL2013}
{Kunz}, M.~W. \& {Lesur}, G. 2013, \mnras, 434, 2295

\bibitem[{{R{\"u}diger} \& {Kitchatinov}(2005)}]{RK2005}
{R{\"u}diger}, G. \& {Kitchatinov}, L.~L. 2005, \aap, 434, 629

\bibitem[{{Sano} \& {Stone}(2002)}]{SS2002}
{Sano}, T. \& {Stone}, J.~M. 2002, \apj, 570, 314

\bibitem[{{Stone} {et~al.}(2000){Stone}, {Gammie}, {Balbus}, \&
  {Hawley}}]{Stoneetal2000}
{Stone}, J.~M., {Gammie}, C.~F., {Balbus}, S.~A., \& {Hawley}, J.~F. 2000,
  Protostars and Planets IV, 589

\bibitem[{{Symington} {et~al.}(2005){Symington}, {Harries}, {Kurosawa}, \&
  {Naylor}}]{Symington2005}
{Symington}, N.~H., {Harries}, T.~J., {Kurosawa}, R., \& {Naylor}, T. 2005,
  \mnras, 358, 977

\bibitem[{{Terquem} \& {Papaloizou}(1996)}]{TP1996}
{Terquem}, C. \& {Papaloizou}, J.~C.~B. 1996, \mnras, 279, 767

\bibitem[{{Valenti} \& {Johns-Krull}(2004)}]{ValentiJK2004}
{Valenti}, J.~A. \& {Johns-Krull}, C.~M. 2004, \apss, 292, 619

\bibitem[{{Wardle}(1999)}]{Wardle1999}
{Wardle}, M. 1999, \mnras, 307, 849

\bibitem[{{Wardle} \& {Salmeron}(2012)}]{WS2012}
{Wardle}, M. \& {Salmeron}, R. 2012, \mnras, 422, 2737

\end{thebibliography}

%\appendix

\end{document}